\documentclass[acus]{JAC2001}
\usepackage{graphicx}
\newcommand{\ud}{\mathrm{d}}
\begin{document}

\onecolumn 
\thispagestyle{empty}

\begin{center}
\begin{tabular}{p{130mm}}

\begin{center}
{\bf\Large
FROM LOCALIZATION TO STOCHASTICS} \\
\vspace{5mm}

{\bf\Large IN BBGKY COLLECTIVE DYNAMICS}

\vspace{1cm}

{\bf\Large Antonina N. Fedorova, Michael G. Zeitlin}\\

\vspace{1cm}

{\bf\large\it
IPME RAS, St.~Petersburg, 
V.O. Bolshoj pr., 61, 199178, Russia}\\
{\bf\large\it e-mail: zeitlin@math.ipme.ru}\\
{\bf\large\it http://www.ipme.ru/zeitlin.html}\\
{\bf\large\it http://www.ipme.nw.ru/zeitlin.html}
\end{center}

\vspace{1cm}

\abstract{ 
Fast and efficient numerical-analytical approach
is proposed for modeling complex collective behaviour
in accelerator/plasma physics models based on BBGKY hierarchy of kinetic equations.
Our calculations are based on                                          
variational and multiresolution approaches in the bases of            
polynomial tensor algebras of generalized coherent states/wavelets. 
We construct the representation for hierarchy of reduced distribution functions
via the multiscale
decomposition in
high-localized eigenmodes.
Numerical modeling shows the creation of
different internal coherent
structures from localized modes, which are related to stable/unstable type
of behaviour and
corresponding pattern (waveletons) formation.
}

\vspace{60mm}

\begin{center}
{\large Presented at the Eighth European Particle Accelerator Conference} \\
{\large EPAC'02} \\
{\large Paris, France,  June 3-7, 2002}
\end{center}
\end{tabular}
\end{center}
\newpage

\title{FROM LOCALIZATION TO STOCHASTICS IN BBGKY COLLECTIVE DYNAMICS}
\author{Antonina N. Fedorova, Michael G. Zeitlin \\
IPME RAS, St.~Petersburg, 
V.O. Bolshoj pr., 61, 199178, Russia
\thanks{e-mail: zeitlin@math.ipme.ru}\thanks{http://www.ipme.ru/zeitlin.html;
http://www.ipme.nw.ru/zeitlin.html}}

\maketitle

\begin{abstract}
Fast and efficient numerical-analytical approach
is proposed for modeling complex collective behaviour
in accelerator/plasma physics models based on BBGKY hierarchy of kinetic equations.
Our calculations are based on                                          
variational and multiresolution approaches in the bases of            
polynomial tensor algebras of generalized coherent states/wavelets. 
We construct the representation for hierarchy of reduced distribution functions
via the multiscale
decomposition in
high-localized eigenmodes.
Numerical modeling shows the creation of
different internal coherent
structures from localized modes, which are related to stable/unstable type
of behaviour and
corresponding pattern (waveletons) formation.
\end{abstract}

\section{INTRODUCTION}

The kinetic theory describes a lot of phenomena in beam/plasma physics
which cannot be understood on the thermodynamic or/and fluid models level.
We mean first of all (local) fluctuations from equilibrium state and 
collective/relaxation phenomena.
It is well-known that only kinetic approach can describe Landau damping, 
intra-beam scattering, while Schottky noise and associated cooling 
technique depend on the understanding of spectrum of local fluctuations
of the beam charge density [1].  
In this paper we consider the applications of a new nu\-me\-ri\-cal\-analytical 
technique based on wavelet analysis approach for 
calculations related to description of complex collective behaviour in the framework of
general BBGKY hierarchy.  
The rational type of
nonlinearities allows us to use our results from 
[2]-[15], which are based on the application of wavelet analysis technique and 
variational formulation of initial nonlinear problems.
Wavelet analysis is a set of mathematical
methods which give us a possibility to work with well-localized bases in
functional spaces and provide maximum sparse forms  for the general 
type of operators (differential,
integral, pseudodifferential) in such bases. 
It provides the best possible rates of convergence and minimal complexity 
of algorithms inside and 
as a result saves CPU time and HDD space.
In part 2  set-up for kinetic BBGKY hierarchy is described.
In part 3 we present explicit analytical construction for solutions of
hierarchy of equations from part 2, which is based on tensor algebra extensions of multiresolution
representation and variational formulation.
We give explicit representation for hierarchy of n-particle reduced distribution functions 
in the base of
high-localized generalized coherent (regarding underlying affine group) 
states given by polynomial tensor algebra of wavelets, which 
takes into account
contributions from all underlying hidden multiscales
from the coarsest scale of resolution to the finest one to
provide full information about stochastic dynamical process.
So, our approach resembles Bogolubov and related approaches 
but we don't use any perturbation technique (like virial expansion)
or linearization procedures.
Numerical modeling shows the creation of
different internal (coherent)
structures from localized modes, which are related to stable (equilibrium) or unstable type
of behaviour and
corresponding pattern (waveletons) formation.

\section{BBGKY HIERARCHY}

Let M be the phase space of ensemble of N particles ($ {\rm dim}M=6N$)
with coordinates
$x_i=(q_i,p_i), \quad i=1,...,N, \quad
q_i=(q^1_i,q^2_i,q^3_i)\in R^3,\quad
p_i=(p^1_i,p^2_i,p^3_i)\in R^3,\quad
q=(q_1,\dots,q_N)\in R^{3N}$.
Individual and collective measures are: 
\begin{eqnarray}
\mu_i=\ud x_i=\ud q_ip_i,\quad \mu=\prod^N_{i=1}\mu_i
\end{eqnarray}
Distribution function
$D_N(x_1,\dots,x_N;t)$
satisfies 
Liouville equation of motion for ensemble with Hamiltonian $H_N$ :
\begin{eqnarray}
\frac{\partial D_N}{\partial t}=\{H_N,D_N\}
\end{eqnarray}
and normalization constraint
\begin{eqnarray}
\int D_N(x_1,\dots,x_N;t)\ud\mu=1
\end{eqnarray}
where Poisson brackets are:
\begin{eqnarray}
\{H_N,D_N\}=\sum^N_{i=1}\Big(\frac{\partial H_N}{\partial q_i}
\frac{\partial D_N}{\partial p_i} - \frac{\partial H_N}{\partial p_i}
\frac{\partial D_N}{\partial q_i}\Big)
\end{eqnarray}
Our constructions can be applied to the following general Hamiltonians:
\begin{eqnarray}
H_N=\sum^N_{i=1}\Big(\frac{p^2_i}{2m}+U_i(q)\Big)+
\sum_{1\leq i\leq j\leq N}U_{ij}(q_i,q_j)  
\end{eqnarray}
where potentials 
$U_i(q)=U_i(q_1,\dots,q_N)$ and $U_{ij}(q_i,q_j)$
are not more than rational functions on coordinates.
Let $L_s$ and $L_{ij}$ be the Liouvillean operators (vector fields)
\begin{eqnarray}
L_s=\sum^s_{j=1}\Big(\frac{p_j}{m}\frac{\partial}{\partial q_j}-
\frac{\partial u_j}{\partial q}\frac{\partial}{\partial p_j}\Big)-\sum_{1\leq i\leq j\leq s}L_{ij}
\end{eqnarray}
\begin{eqnarray}
L_{ij}=\frac{\partial U_{ij}}{\partial q_i}\frac{\partial}{\partial p_i}+
\frac{\partial U_{ij}}{\partial q_j}\frac{\partial}{\partial p_j}
\end{eqnarray}
For s=N we have the following representation for Liouvillean vector field
\begin{eqnarray}
L_N=\{H_N,\cdot \}
\end{eqnarray}
and the corresponding ensemble equation of motion:
\begin{eqnarray}
\frac{\partial D_N}{\partial t}+L_ND_N=0
\end{eqnarray}
$L_N$ is 
self-adjoint operator regarding standard pairing on the set of phase space functions.
Let
\begin{eqnarray}
F_N(x_1,\dots,x_N;t)=\sum_{S_N}D_N(x_1,\dots,x_N;t)
\end{eqnarray}
be the N-particle distribution function ($S_N$ is permutation group of N elements). 
Then we have the hierarchy of reduced distribution functions ($V^s$ is the
corresponding normalized volume factor) 
\begin{eqnarray}
&&F_s(x_1,\dots,x_s;t)=\\
&&V^s\int D_N(x_1,\dots,x_N;t)\prod_{s+1\leq i\leq N}\mu_i\nonumber
\end{eqnarray}
After standard manipulations we arrived to BBGKY hierarchy [1]:
\begin{eqnarray}
\frac{\partial F_s}{\partial t}+L_sF_s=\frac{1}{\upsilon}\int\ud\mu_{s+1}
\sum^s_{i=1}L_{i,s+1}F_{s+1}
\end{eqnarray}
It should be noted that we may apply our approach even to more general formulation than
(12). Some particular case is considered in [16]. 

\section{MULTISCALE ANALYSIS}

The infinite hierarchy of distribution functions satisfying system (12)
in the thermodynamical limit is:
\begin{eqnarray}
&&F=\{F_0,F_1(x_1;t),F_2(x_1,x_2;t),\dots,\\
&&F_N(x_1,\dots,x_N;t),\dots\}\nonumber
\end{eqnarray}
where
$F_p(x_1,\dots, x_p;t)\in H^p$,
$H^0=R,\quad H^p=L^2(R^{6p})$ (or any different proper functional space), $F\in$
$H^\infty=H^0\oplus H^1\oplus\dots\oplus H^p\oplus\dots$
with the natural Fock-space like norm (of course, we keep in mind 
the positivity of the full measure):
\begin{eqnarray}
(F,F)=F^2_0+\sum_{i}\int F^2_i(x_1,\dots,x_i;t)\prod^i_{\ell=1}\mu_\ell
\end{eqnarray}
First of all we consider $F=F(t)$ as function on time variable only,
$F\in L^2(R)$, via
multiresolution decomposition which naturally and efficiently introduces 
the infinite sequence of underlying hidden scales into the game [17].
Because affine
group of translations and dilations is inside the approach, this
method resembles the action of a microscope. We have contribution to
final result from each scale of resolution from the whole
infinite scale of spaces. Let the closed subspace
$V_j (j\in {\bf Z})$ correspond to  level j of resolution, or to scale j.
We consider  a multiresolution analysis of $L^2(R)$
(of course, we may consider any different functional space)
which is a sequence of increasing closed subspaces $V_j$:
$
...V_{-2}\subset V_{-1}\subset V_0\subset V_{1}\subset V_{2}\subset ...
$
satisfying the following properties:
let $W_j$ be the orthonormal complement of $V_j$ with respect to $V_{j+1}$: 
$
V_{j+1}=V_j\bigoplus W_j
$
then we have the following decomposition:
\begin{eqnarray}
\{F(t)\}=\bigoplus_{-\infty<j<\infty} W_j
\end{eqnarray}
or  in case when $V_0$ is the coarsest scale of resolution:
\begin{eqnarray}
\{F(t)\}=\overline{V_0\displaystyle\bigoplus^\infty_{j=0} W_j},
\end{eqnarray}
Subgroup of translations generates basis for fixed scale number:
$
{\rm span}_{k\in Z}\{2^{j/2}\Psi(2^jt-k)\}=W_j.
$
The whole basis is generated by action of full affine group:
\begin{eqnarray}
&&{\rm span}_{k\in Z, j\in Z}\{2^{j/2}\Psi(2^jt-k)\}=\\ 
&&{\rm span}_{k,j\in Z}\{\Psi_{j,k}\}
=\{F(t)\}\nonumber
\end{eqnarray}
Let sequence $\{V_j^t\}, V_j^t\subset L^2(R)$ correspond to multiresolution analysis on time axis, 
$\{V_j^{x_i}\}$ correspond to multiresolution analysis for coordinate $x_i$,
then
\begin{equation}
V_j^{n+1}=V^{x_1}_j\otimes\dots\otimes V^{x_n}_j\otimes  V^t_j
\end{equation}
corresponds to multiresolution analysis for n-particle distribution fuction 
$F_n(x_1,\dots,x_n;t)$.
E.g., for $n=2$:
{\setlength\arraycolsep{0mm}
\begin{eqnarray}
&&V^2_0=\{f:f(x_1,x_2)=\\
&&\sum_{k_1,k_2}a_{k_1,k_2}\phi^2(x_1-k_1,x_2-k_2),\ 
a_{k_1,k_2}\in\ell^2(Z^2)\},\nonumber
\end{eqnarray}}
where 
$
\phi^2(x_1,x_2)=\phi^1(x_1)\phi^2(x_2)=\phi^1\otimes\phi^2(x_1,x_2),
$
and $\phi^i(x_i)\equiv\phi(x_i)$ form a multiresolution basis corresponding to
$\{V_j^{x_i}\}$.
If $\{\phi^1(x_1-\ell)\},\ \ell\in Z$ form an orthonormal set, then 
$\phi^2(x_1-k_1, x_2-k_2)$ form an orthonormal basis for $V^2_0$.
Action of affine group provides us by multiresolution representation of
$L^2(R^2)$. After introducing detail spaces $W^2_j$, we have, e.g. 
$
V^2_1=V^2_0\oplus W^2_0.
$
Then
3-component basis for $W^2_0$ is generated by translations of three functions 
\begin{eqnarray}
&&\Psi^2_1=\phi^1(x_1)\otimes\Psi^2(x_2),\ \Psi^2_2=\Psi^1(x_1)\otimes\phi^2(x_2), \nonumber\\
&&\Psi^2_3=\Psi^1(x_1)\otimes\Psi^2(x_2)
\end{eqnarray}
Also, we may use the rectangle lattice of scales and one-dimentional wavelet
decomposition :
$$
f(x_1,x_2)=\sum_{i,\ell;j,k}<f,\Psi_{i,\ell}\otimes\Psi_{j,k}>
\Psi_{j,\ell}\otimes\Psi_{j,k}(x_1,x_2)
$$
where bases functions $\Psi_{i,\ell}\otimes\Psi_{j,k}$ depend on
two scales $2^{-i}$ and $2^{-j}$.
After constructing multidimension bases we apply one of variational procedures from
[2]-[16].
As a result the solution of equations (12) has the 
following mul\-ti\-sca\-le\-/mul\-ti\-re\-so\-lu\-ti\-on decomposition via 
nonlinear high\--lo\-ca\-li\-zed eigenmodes 
{\setlength\arraycolsep{0pt}
\begin{eqnarray}
&&F(t,x_1,x_2,\dots)=\sum_{(i,j)\in Z^2}a_{ij}U^i\otimes V^j(t,x_1,x_2,\dots)\nonumber\\
&&V^j(t)=V_N^{j,slow}(t)+\sum_{l\geq N}V^j_l(\omega_lt), \quad \omega_l\sim 2^l \\
&&U^i(x_s)=U_M^{i,slow}(x_s)+\sum_{m\geq M}U^i_m(k^{s}_mx_s), \quad k^{s}_m\sim 2^m,
 \nonumber
\end{eqnarray}}
which corresponds to the full multiresolution expansion in all underlying time/space 
scales.
\begin{figure}[htb]
\centering
\includegraphics*[width=60mm]{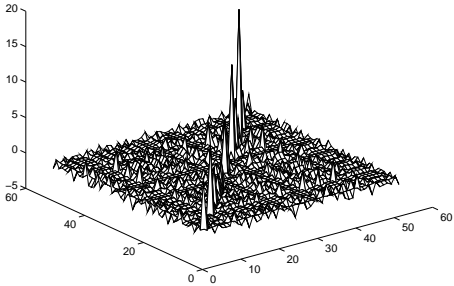}
\caption{6-eigenmodes representation.}
\end{figure} 
Formulae (21) give us expansion into the slow part $\Psi_{N,M}^{slow}$
and fast oscillating parts for arbitrary N, M.  So, we may move
from coarse scales of resolution to the 
finest one for obtaining more detailed information about our dynamical process.
The first terms in the RHS of formulae (21) correspond on the global level
of function space decomposition to  resolution space and the second ones
to detail space. In this way we give contribution to our full solution
from each scale of resolution or each time/space scale or from each nonlinear eigenmode.
It should be noted that such representations 
give the best possible localization
properties in the corresponding (phase)space/time coordinates. 
In contrast with different approaches formulae (21) do not use perturbation
technique or linearization procedures.
Numerical calculations are based on compactly supported
wavelets and related wavelet families and on evaluation of the accuracy 
regarding norm (14):
\begin{equation}
\|F^{N+1}-F^{N}\|\leq\varepsilon
\end{equation}
Fig.~1 demonstrates 6-scale/eigenmodes (waveletons) construction for solution of
equations like (12).
So, by using wavelet bases with their good (phase) space/time      
localization properties we can construct high-localized waveleton structures in      
spa\-ti\-al\-ly\--ex\-te\-nd\-ed stochastic systems with collective behaviour.

\end{document}